**20-125 MeV/nuc COSMIC RAY CARBON NUCLEI INTENSITIES**

**BETWEEN 2004-2010 IN SOLAR CYCLE #23 AS MEASURED NEAR THE EARTH,**

**AT VOYAGER 2 AND ALSO IN THE HELIOSHEATH AT VOYAGER 1 -**

**MODULATION IN A TWO ZONE HELIOSPEHRE**

**W.R Webber[1], A.C. Cummings[2], E.C. Stone[2], F.B. McDonald[3], R.A. Mewaldt[2],**

**R. Leske[2], M. Wiedenbeck[2], P.R. Higbie[4], and B. Heikkila[5]**

1. New Mexico State University, Department of Astronomy, Las Cruces, NM, 88003, USA

2. California Institute of Technology, Space Radiation Laboratory, Pasadena, CA, 91125, USA

3. University of Maryland, Institute of Physical Science and Technology, College Park, MD, 20742, USA

4. New Mexico State University, Physics Department, Las Cruces, NM, 88003, USA

5. NASA/Goddard Space Flight Center, Greenbelt, MD  20771, USA

**Abstract**

The recovery of cosmic ray Carbon nuclei of energy ~20-125 MeV/nuc in solar cycle #23 from 2004 to 2010 has been followed at three locations, near the Earth using ACE data and at V2 between 74-92 AU and also at V1 beyond the heliospheric termination shock at between 91-113 AU. To describe the observed intensity changes and to predict the absolute intensities measured at all three locations we have used a simple spherically symmetric (no drift) two-zone heliospheric transport model with specific values for the diffusion coefficient in both the inner and outer zones. The diffusion coefficient in the outer zone is determined to be ~5-10 times smaller than that in the inner zone out to 90 AU. For both V1 and V2 the calculated C nuclei intensities agree within an average of ± 10% with the observed intensities. Because of this agreement between V1 and V2 observations and predictions there is no need to invoke an asymmetrical squashed heliosphere or other effects to explain the V2 intensities relative to V1 as is the case for He nuclei. The combination of the diffusion parameters used in this model and the interstellar spectrum give an unusually low overall solar modulation parameter $\phi = 250$ MV to describe the Carbon intensities observed at the Earth in 2009. At all times both the observed and calculated spectra are very closely ~ $E^{1.0}$ as would be expected in the adiabatic energy loss regime of solar modulation.

**Introduction**

The intensity recovery of lower energy galactic cosmic rays at the Earth in the solar 11-year cycle #23 between 2004-2010 and the unusually high intensities observed in 2009 is well documented using spacecraft data (e.g., McDonald, Webber and Reames, 2010; Mewaldt, et al., 2010). The cosmic ray recovery in this cycle started in early 2004 at the Earth after the large "Halloween" events in October-November, 2003, and has been observed by neutron monitors and various spacecraft near the Earth including ACE, IMP and others. This recovery was observed by V2 and V1 to begin in the outer heliosphere in late 2004 after the Halloween event had propagated out to their respective locations at 76 and 93 AU (McDonald, et al., 2006). At the end of 2004 V1 crossed the HTS at 94 AU and has continued to move outward so that by 2010.5 it was at ~114 AU, perhaps ~30 AU or more beyond the current HTS location, estimated to be between 80-85 AU (Webber and Intriligator, 2011). Thus V1 has spent essentially the entire recovery cycle beyond the HTS in the heliosheath region where the solar wind parameters are measurably different from those in the inner heliosphere. V2 remained in the "inner" part of the heliosphere until 2007.66 when at ~84 AU it also crossed the HTS.

At about 2010.0 the cosmic ray Carbon nuclei intensity at the Earth reached its maximum (Mewaldt, et al., 2010). At V1 the intensity of Carbon nuclei continues to increase as of 2010.5 whereas at V2 it reached a maximum in early 2009, then decreased, but after 2010.5 began a rapid increase. At the Earth the Carbon intensities reached levels ~25% higher than those observed during the previous 11-year intensity maximum in 1997-98 (McDonald, Webber and Reames, 2010; Mewaldt, et al., 2010). At V1 the cosmic ray Carbon intensities are at the highest levels yet observed and at energies ~100 MeV/nuc at 2010.5 are within ~20% of the estimated LIS intensities for Carbon nuclei (see Webber and Higbie, 2009; George, et al., 2009).

It is the purpose of this paper to compare the Carbon nuclei intensities between 20-125 MeV/nuc observed at the Earth and those observed at V1 and V2 during this time period, within the framework of simple modulation models, with the objective of understanding better the global characteristics of the solar 11-year modulation cycle, including, particularly, the modulation effects beyond the HTS in the heliosheath.

This is the third of several articles dealing with the recovery of cosmic ray intensities at V1, V2 and the Earth during this extended time period. The initial article considered He nuclei between 150-250 MeV/nuc (Webber, et al., 2011a). The second article studies the H nuclei from

69  150-250 MeV (Webber, et al., 2011b).  Each of these individual nuclei gives its own specific

70  information on the overall radial extent of the heliosphere modulation process and the required

71  interstellar spectrum of the galactic particles involved.  These studies of Carbon nuclei, which

72  simultaneously involve the inner and outer heliosphere, are now possible because of ACE

73  measurements at the Earth covering an entire solar 11 year cycle from 1997 to 2010, and the V1

74  (V2) measurements from the outer heliosphere including the heliosheath from 2005 to 2010.

75      The energies involved in the Carbon nuclei study are in the low energy part of the

76  spectrum below ~125 MeV/nuc where the energy spectrum, dj/dE, has an $E^{1.0}$ dependence, the so

77  called adiabatic regime in which adiabatically cooled higher energy particles populate the

78  spectrum.   This is in contrast to the He nuclei and H nuclei studies at ~200 MeV/nuc noted

79  above, which are above the peak in the differential energy spectrum and in an energy region

80  where "diffusive" modulation effects are important.

81  **Observations at the Earth and at V1 and V2**

82      In Figures 1A and 1B we show the time history of ~56-125 and 20-56 MeV/nuc Carbon

83  nuclei from 2004 to 2010.5.  The Voyager data is smoothed by taking 5 interval 26 day moving

84  averages.  The data at the Earth is from the ACE/CRIS experiment from Mewaldt, et al., 2010

85  (also http://www.srl.caltech.edu/ACE).   For the energy interval 94-125 MeV/nuc we use the O

86  nuclei intensities at V1 and V2 as a proxy for the C nuclei since this energy interval is beyond

87  the range for C.  At this energy it has been shown (e.g., the ACE web-site) that the C/O ratio is =

88  1.05 ± 0.05 for a wide range of modulation levels, therefore this is a strong proxy.  As a result we

89  can use energy intervals for C nuclei on V1 and V2 that almost exactly match those available

90  from ACE from 20 to 125 MeV/nuc.

91      At the beginning of the recovery time period the intensity of 56-125 MeV/nuc Carbon at

92  V1 was ~3-4 times that at the Earth.  This is a measure of the overall interplanetary gradient at

93  this energy between 1 and ~94 AU, the location of V1 at that time.  By 2010.5 this intensity ratio

94  is reduced to ~2 implying that the intensity changes between 2004 and 2010.5 at the Earth are

95  greater than those at V1.  For 20-56 MeV/nuc particles these ratios at 2004 and 2010.5 are ~3

96  and 2 respectively.  These changing intensity ratios at both energies are shown in Figure 2.

97      In Figure 3 we show the 56-125 MeV/nuc data at the Earth superimposed on the data at

98  V1 (with different intensity scales), with the data at Earth delayed to account for the solar wind

99  propagation time from the Earth to V1.  This delay time is varied from 0.5 to 1.5 years in 26 day

100 increments and the correlation coefficient rapidly improves, reaching a maximum value 0.923 for
101 time delays between 0.86 and 0.93 years. This correspondence of time histories is remarkable
102 considering the 100 AU difference in the radial location of the spacecraft.

103     This correlation throughout the heliosphere is also evident in Figures 4A and 4B which
104 shows the intensities at each energy at V1 and V2 vs. those at the Earth, with a delay ~0.89 yrs.

105     We seek to fit the data in Figures 4A and 4B and to interpret it using a simple global
106 modulation model in conjunction with a realistic interstellar (IS) C spectrum. The goal of this
107 calculation is to predict the absolute intensities at all three locations and also the changing ratios
108 of intensities at V1 beyond the HTS, at V2 mainly just inside the HTS, and the intensities at the
109 Earth vs. time as given by Figures 1A and 1B and also Figures 4A and 4B. In addition the
110 calculation should predict the average slopes of the regression lines between V1 and V2 and the
111 Earth data at each location and at each energy as plotted in Figures 4A and 4B.

112 **The Cosmic Ray Transport Equation in the Heliosphere**

113     Here we use a simple spherically symmetric quasisteady state no-drift transport model for
114 cosmic rays in the heliosphere. While this simplified model obviously cannot fit all types of
115 observations it does provide a useful insight into the inner heliospheric/outer heliospheric
116 modulation and helps to determine which aspects of this modulation need more sophisticated
117 models for their explanation. The numerical model was originally provided to us by Moraal
118 (2003, private communications) and is similar to the model described originally in Reinecke,
119 Moraal and McDonald, 1993, and in Caballero-Lopez and Moraal, 2004, and also to the
120 spherically symmetric transport model described by Jokipii, Kota and Merenyi, 1993 (Figure 3
121 of that paper). The basic transport equation is (Gleeson and Urch, 1971);

122
$$\frac{\partial f}{\partial t} + \nabla \cdot (CVf - K \cdot \nabla f) + \frac{1}{3p^2} \frac{\partial}{\partial p} (p^2 V \cdot \nabla f) = Q$$

123 Here f is the cosmic ray distribution function, p is momentum, V is the solar wind velocity,
124 K(r,p,t) is the diffusion tensor, Q is a source term and C is the so called Compton-Getting
125 coefficient.

126     For spherical symmetry (and considering latitude effects to be unimportant for this
127 calculation) the diffusion tensor becomes a single radial coefficient $K_{rr}$. We assume that this
128 coefficient is separable in the form $K_r(r,P) = \beta\ K_1(P)\ K_2\ (r)$, where the rigidity part, $K_1(P) \equiv K1$
129 and radial part, $K_2(r) \equiv K2$. The rigidity dependence of $K_1(P)$ is assumed to be ~P above a low

130    rigidity limit $P_B$. The units of the coefficient $K_{rr}$ are in terms of the solar wind speed V =

131    $4 \cdot 10^2$ km$\cdot$s$^{-1}$, the distance 1 AU=1.5 x $10^8$ Km, so $K_{rr}$ = $6 \cdot 10^{20}$ cm$^2 \cdot$s$^{-1}$ when K1 = 1.0.

132    Based on our earlier studies using He and H nuclei (Webber, et al., 2011a, b) we consider

133    a two zone heliosphere (e.g., Jokipii, Kota and Merenyi, 1993). In this case the inner zone

134    extends out to 90 AU, the average distance to the HTS. In this inner region V=400 km$\cdot$s$^{-1}$ and

135    the diffusion parameters K1 and K2 are determined in our approach by a fit to the cosmic ray

136    data being compared (the Earth and V2).

137    The outer zone extends from 90 AU to ~120 AU, the approximate distance to the

138    heliopause (HP) or an equivalent "outer boundary" and essentially encompasses the heliosheath.

139    In this region V is taken to be 130 km$\cdot$s$^{-1}$ (from V2 measurements, Richardson, et al., 2008) and

140    the diffusion parameters are K1H and K2H, which are different from those in the inner

141    heliosphere, and again determined by the cosmic ray intensity changes at V1. The distance to the

142    HP and the source spectrum are important in this calculation.

143    For the LIS Carbon spectrum we use the recent spectrum of Webber and Higbie, 2009.

144    This spectrum can be approximated to an accuracy ~few % above ~100 MeV/nuc by

145           Carbon FLIS = $(0.0456/T^{2.76})/ (1.63/T^{0.12}+6.10/T^{1.19}+1.07/T^{2.85}+0.12/T^{4.40})$

146    where T is in GeV/nuc. At the average energies of 92 and 38 MeV/nuc for the energy intervals

147    used for the Carbon data, this equation gives input intensities of 2.05 x $10^{-2}$ and 1.15 x $10^{-2}$

148    p/m$^2 \cdot$sr$\cdot$s$\cdot$MeV/nuc at the boundary of 120 AU. The measured intensities at V1 at 2010.5 are 1.73

149    and 0.75 at these energies (see Figures 1A and 1B).

150    For a simple one zone heliosphere with a boundary at 120 AU, a 1$^{st}$ step is to fit the

151    measured intensities at the Earth to determine the overall modulation. The measured intensities

152    at the Earth are 0.075 and 0.039 in the above units at the two average energies of 92 and 38

153    MeV/nuc in late 2009 (see ACE web site). For K2=0 these intensities at the Earth can be fit with

154    a value of K1 = 160. This value for K1 corresponds to a modulation potential = 248 MV in the

155    equivalent force field approximation where the modulation potential is defined as

156    $$\phi = \int_1^{R_B} \frac{V \, dr}{3K1}$$

157    (see Caballero-Lopez and Moraal, 2004).

158    We note that this modulation potential is much lower than the average value of ~400-500

159    MV observed at previous sunspot minima in the modern era from 1950 (see e.g., Webber and

160 Higbie 2010), in keeping with the unusually high intensities observed at this time (McDonald,
161 Webber and Reames, 2010; Mewaldt, et al., 2010). In fact the low modulation potential that we
162 now find using Carbon nuclei is very similar to the modulation potential of 235 MV obtained by
163 Mewaldt, et al, 2010, using ACE measurements of C and Fe nuclei at the Earth also in late 2009.
164 In their calculations, Mewaldt, et al., 2009, used the George, et al., 2009, LIS Carbon spectrum
165 which differs only slightly from the Webber and Higbie, 2009, LIS spectrum above ~100
166 MeV/nuc.

167 In Figure 6 we show the observed spectrum of C nuclei at V1, V2 and the Earth at two
168 times. One is at the beginning of the time interval of interest, at 2005.0. The second is at the end
169 of the time interval at 2010.5. All of these spectra are consistent with spectral slopes = 1.0 for
170 energies below ~100 MeV/nuc. This Figure also illustrates the reasons why the apparent
171 modulation potential at the Earth is so low in 2010. It is because the IS spectra themselves for C
172 are also much lower than many of the earlier IS spectra that have been used (e.g., see Ngobeni
173 and Potgieter, 2011).

174 For a two zone model with the inner heliosphere boundary at the HTS (taken here to be at
175 90 AU) and the HP at 120 AU, we find that values of K1=175 (max) and 42 (min) and K2 = 0 in
176 the inner heliosphere and values of K1H between 18 (max) and 10 (min), and K2H=0 with
177 V=0.33 in the outer heliosphere along with the IS spectrum given by the equation CARBON-
178 FLIS, we are able to fit the average Carbon data in both energy ranges at the Earth and at V1 and
179 V2 as shown in Figures 4A and 4B and Figures 5A and 5B. A graph of this diffusion coefficient
180 vs. time and rigidity is shown in Figure 5 of Webber, et al., 2011a, b.

181 The predicted V1 "fit lines" in Figures 4A and 4B lie an average of 5% above the data at
182 56-125 MeV/nuc and the fit is equally good for 20-56 MeV/nuc. None of the smoothed data
183 points in Figures 4A and 4B or the spectra in Figure 6 lie more than ±15% from the predicted
184 lines.

185 Note that the fits to the Carbon data is the generally good for V2 in contrast to the
186 calculations for He and H nuclei which require some form of an asymmetrical heliosphere along
187 with other effects to explain the V2 intensities which are at times equal to those at V1 (Webber,
188 et al., 2001a, b). The calculation for Carbon is for a spherical heliosphere with no N-S
189 asymmetry included. The Carbon data reflect the spectrum in the "adiabatic" energy range
190 where the spectra are expected to be ~$E^{1.0}$. Indeed the ratio of the intensities in the 56-125 to 20-

56 MeV/nuc intervals remains essentially constant throughout the entire recovery time period from 2005 to 2010 at all three locations, V1, V2 and the Earth. This behavior is shown in Figure 7 for the 20-56 to 56-125 MeV/nuc ratio at V1. The average value for this ratio between 2004 and 2010 is 0.42±0.02 and is the same for all three locations V1, V2 and the Earth. The ratio of the average energies of the two intervals is 0.41. This intensity ratio remains nearly constant with time although the intensities increase by a factor of ~3 at V1 (and V2) between 2004 and 2010. This ratio also is nearly the same at the Earth and at V1 and V2 even though, as we have shown in Figure 2, the intensities at these locations differ by a factor of 2-4. This constancy implies that over this entire time interval and the large range of "effective" modulation levels and radial distances from the Earth to V1 and V2, the modulation "acts" to first order like an adiabatic modulation process similar to that resulting from a simple "force field" modulation of the type first described by Gleason and Axford, 1968.

Thus, in summary, we have the situation where (1): The magnitude of the diffusion coefficient in the outer zone (heliosheath) is ~5-10 times smaller than that in the inner zone. (2): During the intensity recovery from 2004-2010 the diffusion coefficient in the inner zone increases by a factor ~4 whereas in the outer zone this increase is only a factor ~1.80. (3): The modulation in this lower energy region below ~100 MeV/nuc acts like a simple "force field" modulation where the energy spectra of Carbon nuclei, dj/dE, are ~E throughout the radial range from ~1-110 AU and for all modulation levels.

### Summary and Conclusions

The recovery of the intensity of ~20-125 MeV/nuc cosmic ray Carbon nuclei has been followed between 2004-2010 at the Earth using the ACE spacecraft and also at V1 and V2 in the outer heliosphere and in the case of V1, beyond the HTS. The correlation of the intensity changes at the Earth and V1 in the outer heliosphere (correlation coefficient =0.921), ~100 AU apart in radial distance, is remarkable after accounting for a time delay ~0.9 year due to the solar wind propagation. The relative intensities measured at V1, V2 and at the Earth as well as the slope of the regression lines between the measurements at the Earth and at V1 and V2 place limits on the amount of solar modulation in the inner and outer heliosphere and also on the local IS Carbon spectrum. It is found that the data for Carbon nuclei can be reproduced by a simple two zone heliosphere where the intensity changes are due to changes in the cosmic ray diffusion coefficient K in each zone. In the inner zone, out to the HTS assumed to be at 90 AU, the value of K is quite large but varies by a factor ~4 from the minimum to maximum modulation in this part of the solar 11-year cycle. In the outer zone from ~90-120 AU, essentially in the heliosheath, the value of the diffusion coefficient is much smaller, by a factor ~5-10 and varies by a factor ~1.80 from minimum to maximum modulation. The same set of diffusion coefficients and their variations with rigidity and time that have been used to explain the 150-250 MeV He and H nuclei intensity changes measured at V1, V2 and the Earth during the same time interval (Webber, et al., 2011a, b) have been used in the Carbon analysis.

For V2 these same diffusion coefficients can also explain the intensity variations of Carbon during the recovery between 2004 and 2010. This is in contrast to both the V2 Helium and H nuclei intensity variations during this time period which require an asymmetric squashed heliosphere to explain the H nuclei data plus additional time dependent asymmetric effects to describe the fact that the He nuclei intensities at V2, are at times, equal to those at V1 much further out in the heliosphere.

The Carbon spectrum that is observed throughout this time period as well as the relative modulation at the Earth, V1 and V2 can be described by a simple force field type of modulation, with the spectra at energies from ~20-100 MeV/nuc being described as dj/dE ~E spectra over the entire radial distance range and at all modulation levels (see Figure 6).

The details of the fit to the data beyond the HTS depend on the values of the local interstellar spectrum (LIS) used as an input to the modulation calculation and also the location of the HP or boundary of the modulation region. For the LIS intensities of C nuclei used in this paper, which are lower than many earlier estimates, the data can be well fit for HP distances in the range of 120-130 AU with a simple two zone symmetric heliosphere. The heliosheath region and the lower energy interstellar Carbon spectrum itself below ~125 MeV/nuc will be mapped in more detail as V1 continues to move outward in the heliosphere and the intensities continue to increase towards the LIS value.

**Acknowledgements:** The authors wish to thank the Voyager team (E.C. Stone, P.I.) and the ACE team (Andrew Davis) for making their data available on their web-sites, http://voyager.gsfc.nasa.gov and http://www.srl.caltech.edu/ACE/.

## Figure Captions

**Figure 1A:**  5 x 26-day running average of V1 and V2 and the 27 day average of ACE 56-125 MeV/nuc Carbon nuclei intensities near the Earth from 2004 to 2010.5.  ACE data is delayed by 0.89 year to account for inner-outer heliosphere delay in modulation due to solar wind propagation time.

**Figure 1B:**  5 x 26-day running average of V1 and V2 and the 27 day average of ACE 20-56 MeV/nuc Carbon nuclei intensities near the Earth from 2004 to 2010.5.  ACE data is delayed by 0.89 year to account for inner-outer heliosphere delay in modulation due to solar wind propagation time.

**Figure 2:**  Ratio of V1 to ACE intensities for 56-125 and 20-56 MeV/nuc (in red) Carbon nuclei from 2004 to 2010.5 (Earth data delayed by 0.89 year).

**Figure 3:**  The V1 data in Figure 1 superimposed on the ACE data at the Earth for 56-125 MeV/nuc Carbon delayed by 0.89 year (with different intensity scales on the left and right axis).  This figure shows the high level of correlation between intensity changes at the Earth and in the outer heliosphere during this time period.

**Figure 4A:**  Regression plot of the intensities of 56-125 MeV/nuc Carbon nuclei at V1 and V2 vs. the intensities at the Earth delayed by 0.89 year.

**Figure 4B:**  The same as Figure 4A except for 20-56 MeV/nuc Carbon nuclei.

**Figure 5A:**  Data in Figure 4A superimposed on predictions of a two zone spherically symmetric heliospheric model, $R_B = 120$ AU.

**Figure 5B:**  Data in Figure 4B superimposed on predictions of a two zone spherically symmetric heliospheric model, $R_B = 120$ AU.

**Figure 6:**  The spectra of Carbon nuclei at V1, V2 and the Earth, at 2005.0 and 2010.5.  Note the similarity of all spectra to a dj/dE ~E spectrum at energies less than 100 MeV/nuc.  The estimated IS Carbon spectra from Webber and Higbie, 2009, and George, et al., 2009 are also shown.

**Figure 7:**  Ratios of Carbon intensities in the 20-56 and 56-125 MeV/nuc energy intervals between 1998 and 2010, observed at V1.  The expected value of 0.41 for a dj/dE ~E spectrum is shown as a horizontal solid line.  The values of the ratios at V2 and also at the Earth (ACE) are very similar and are not plotted.

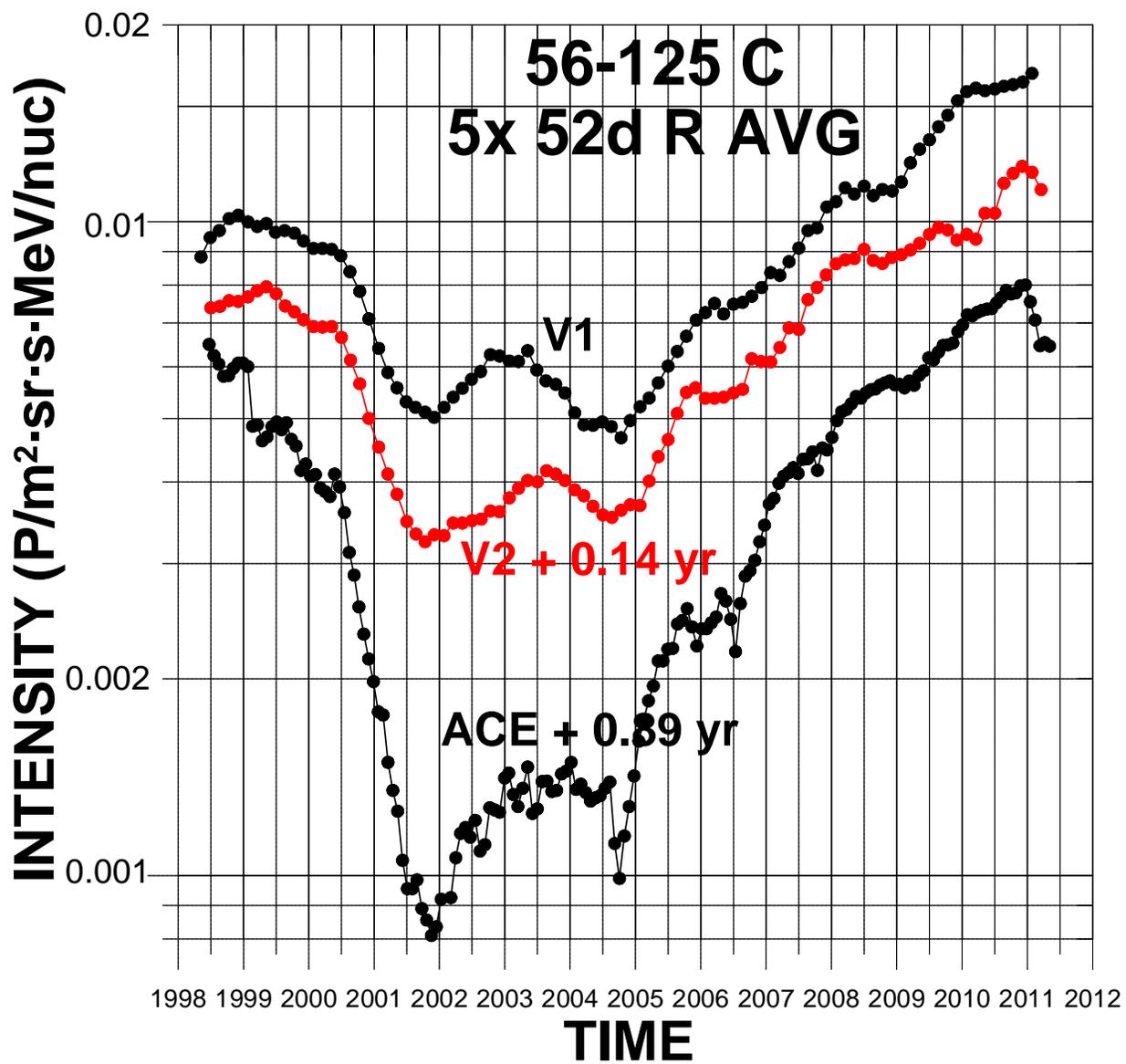

FIGURE 1A

321
322

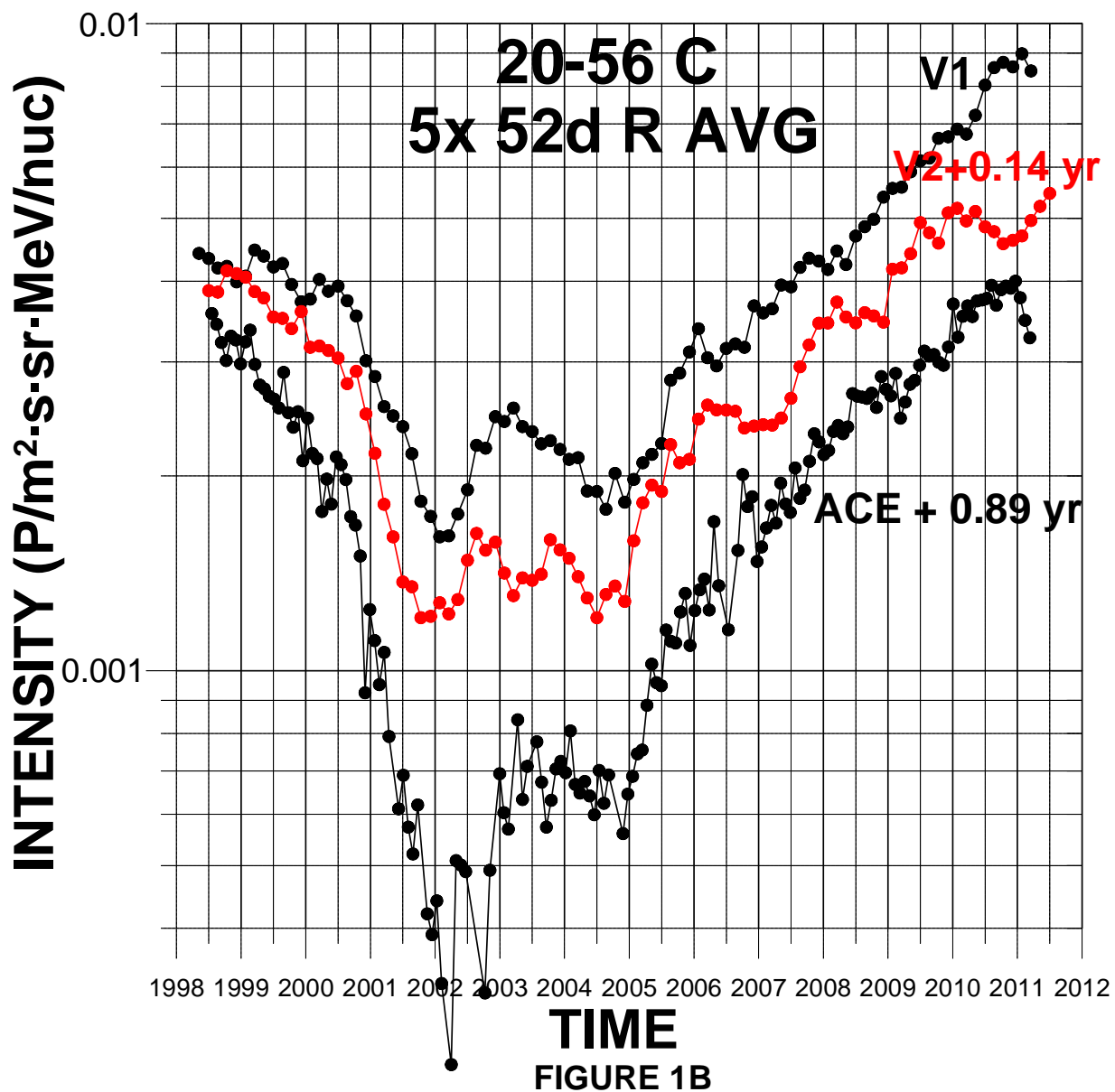

**FIGURE 1B**

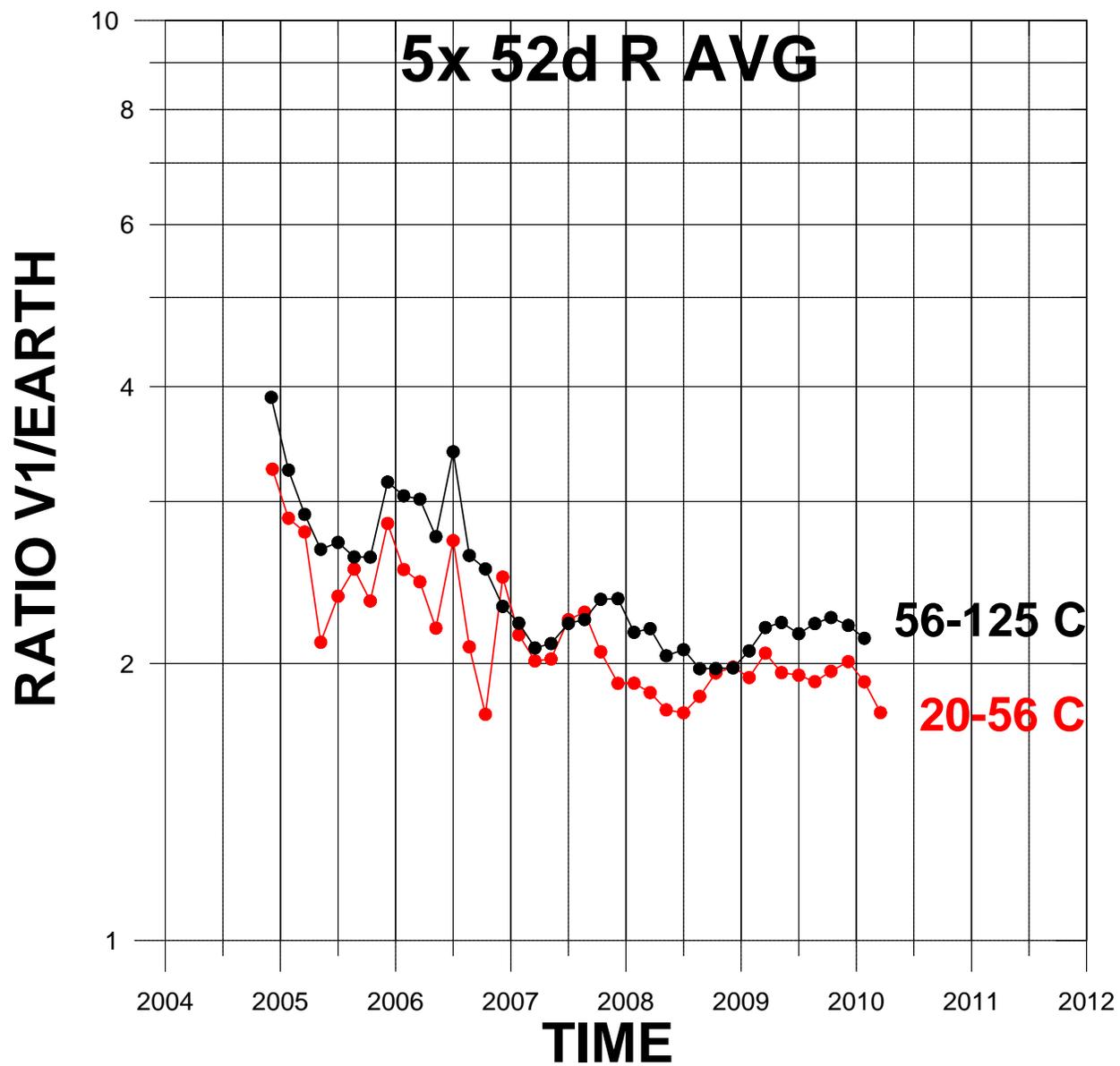

**5x 52d R AVG**

**56-125 C**

**20-56 C**

**TIME**

**FIGURE 2**

—

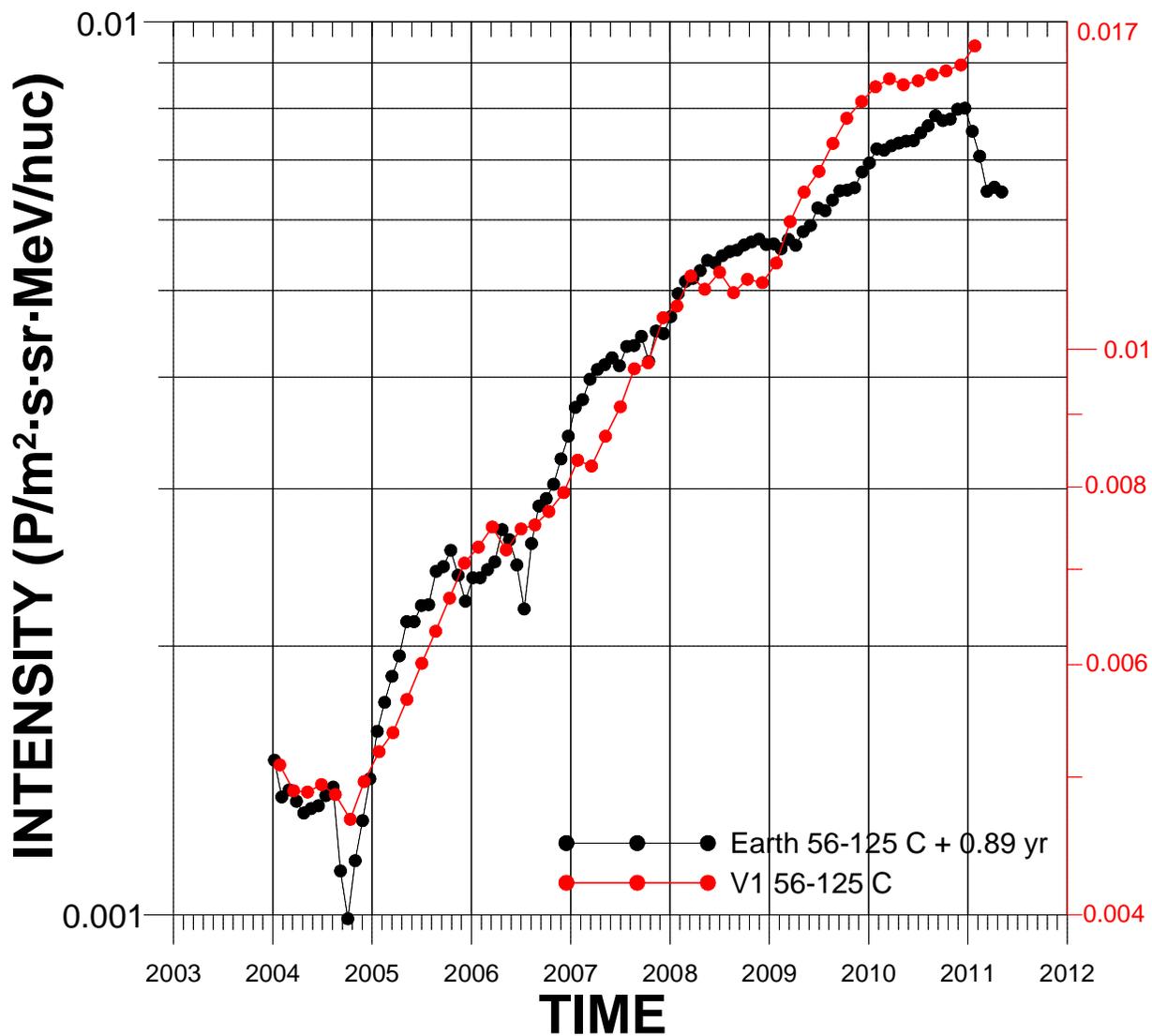

**FIGURE 3**

327
328

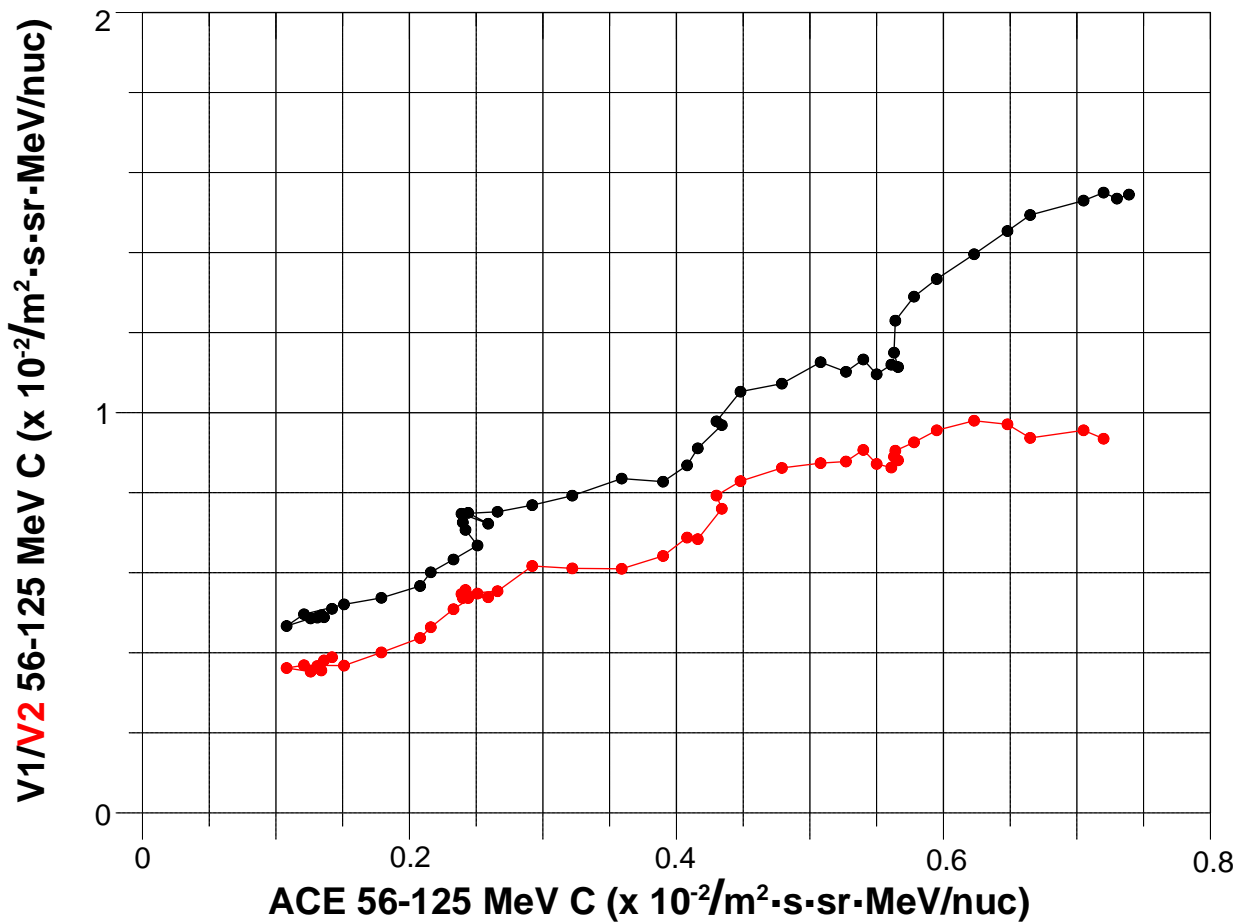

**FIGURE 4A**

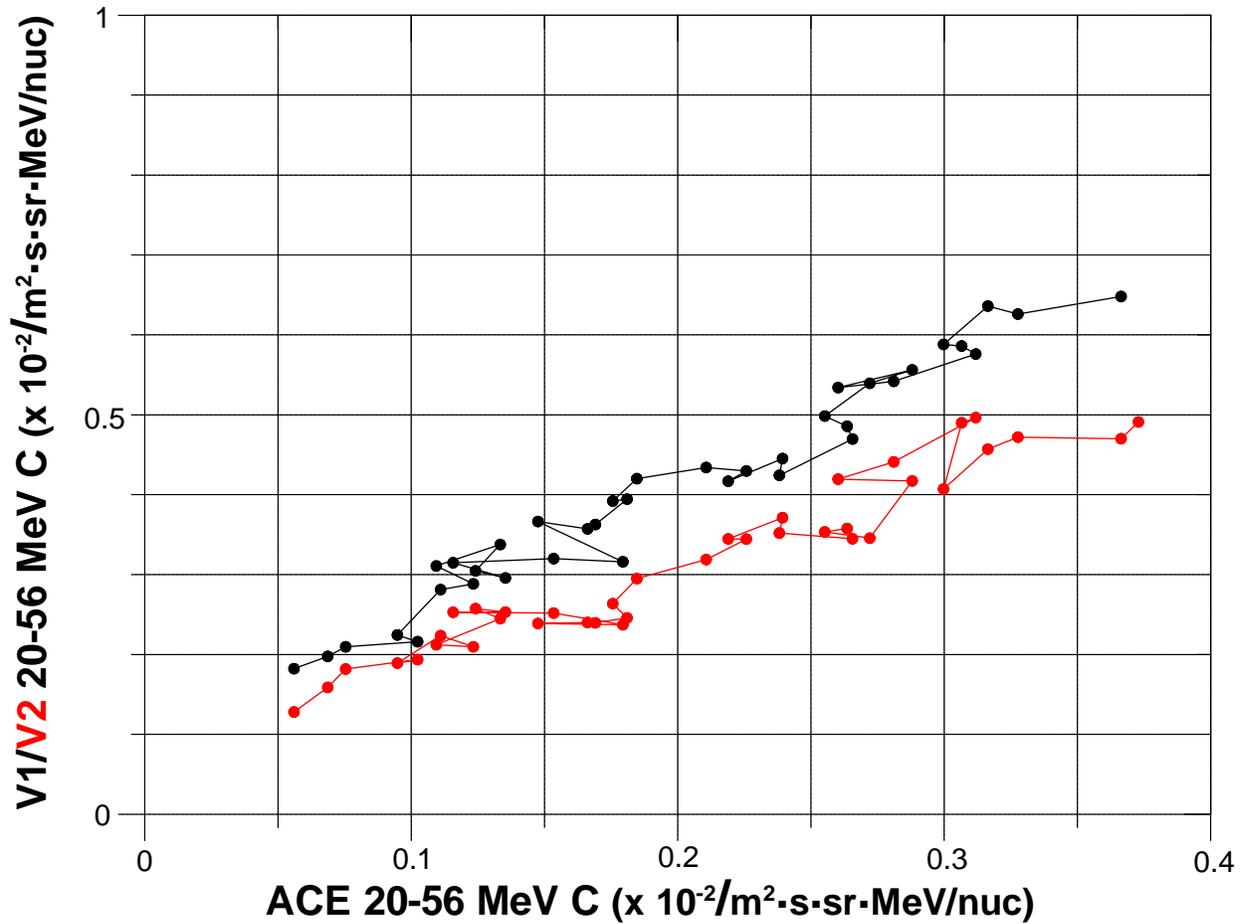

**FIGURE 4B**

331

332

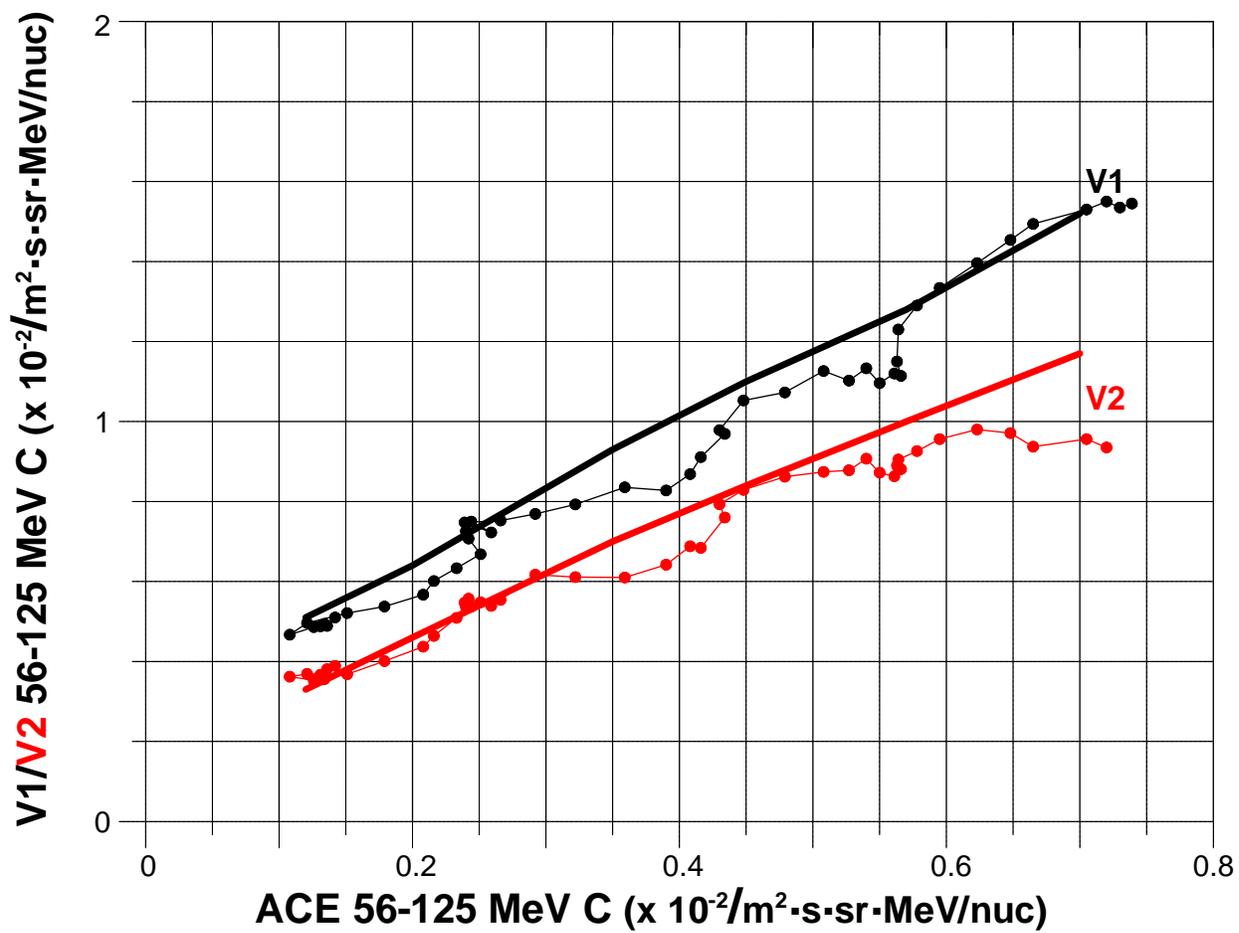

**FIGURE 5A**

333
334

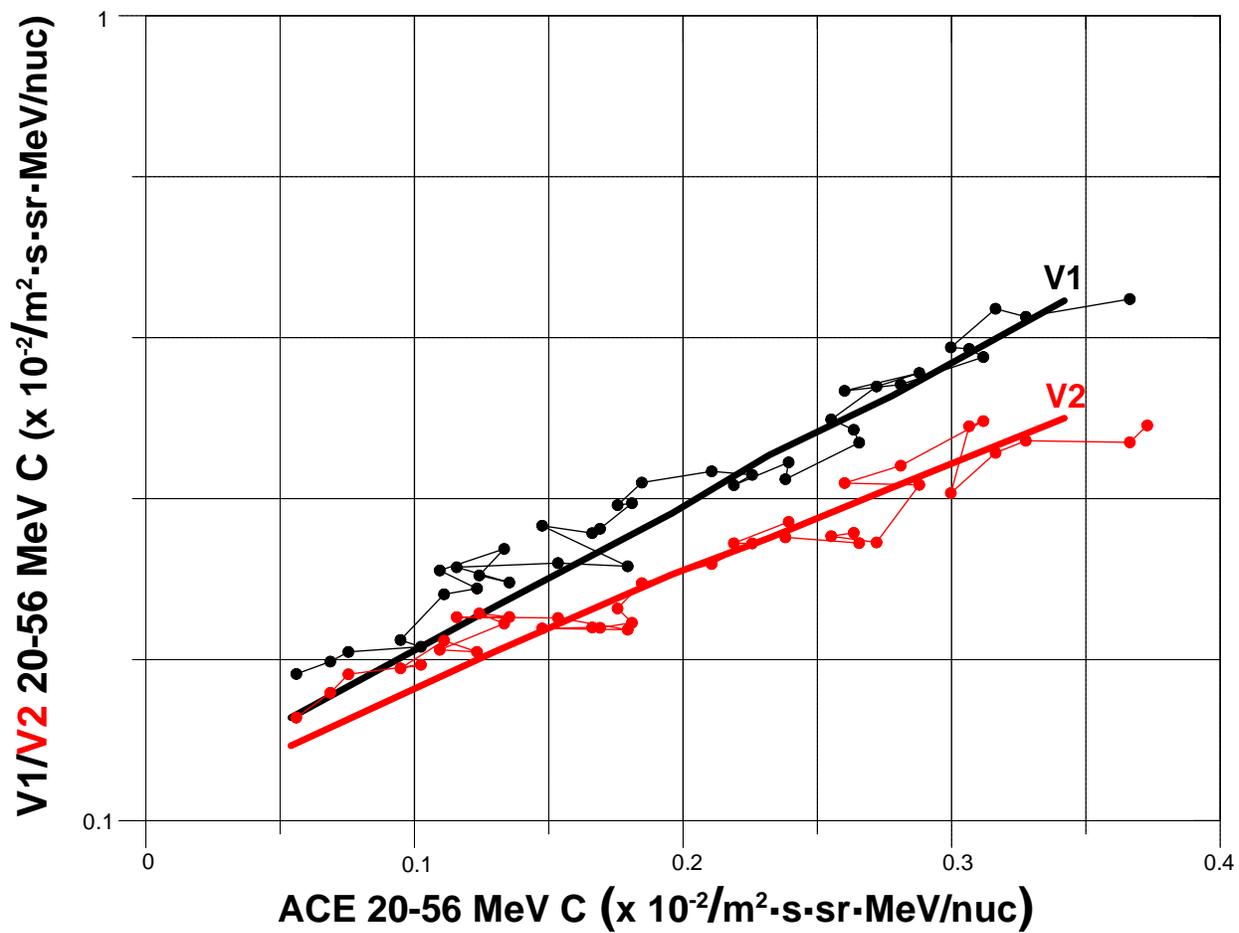

**FIGURE 5B**

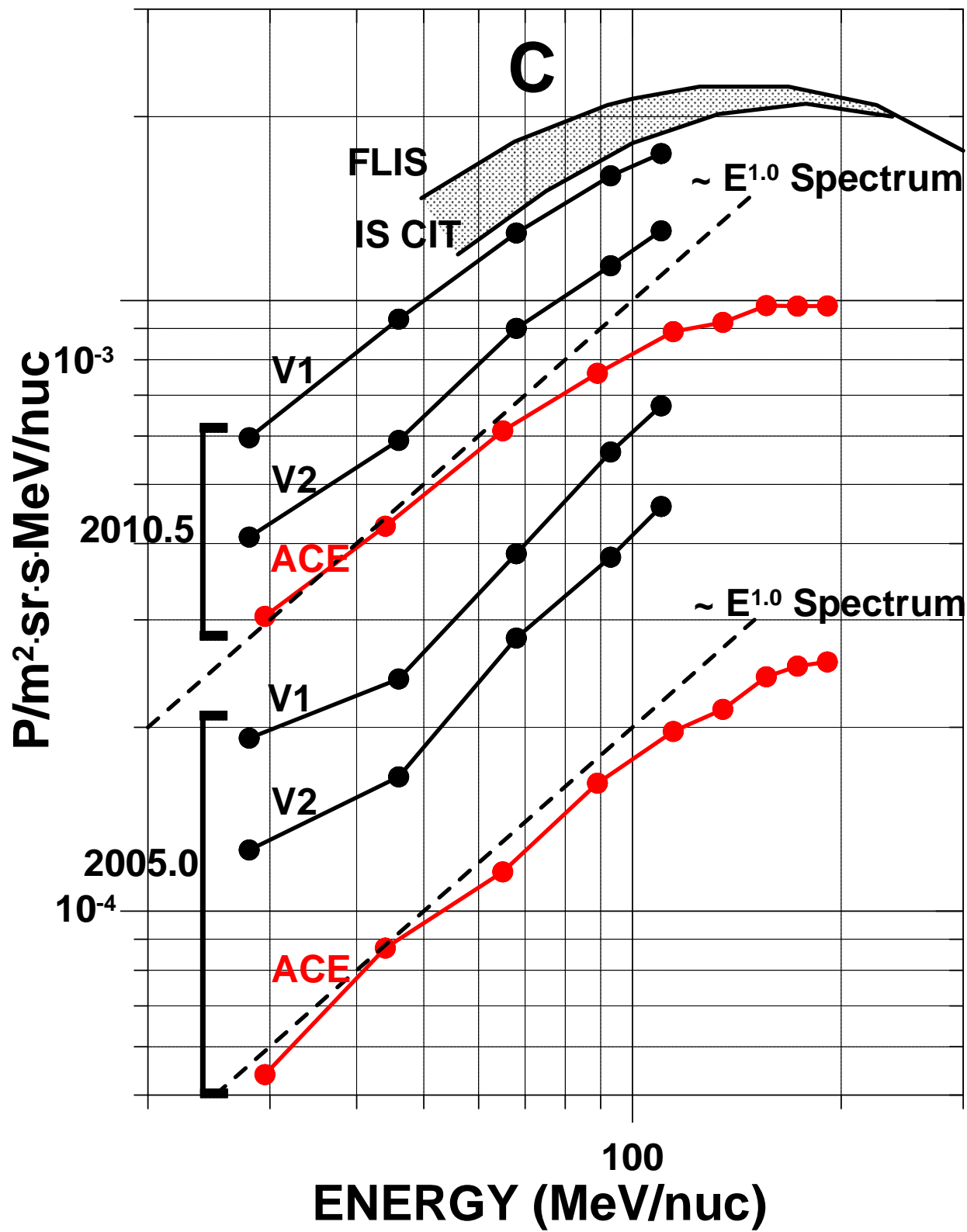

**FIGURE 6**

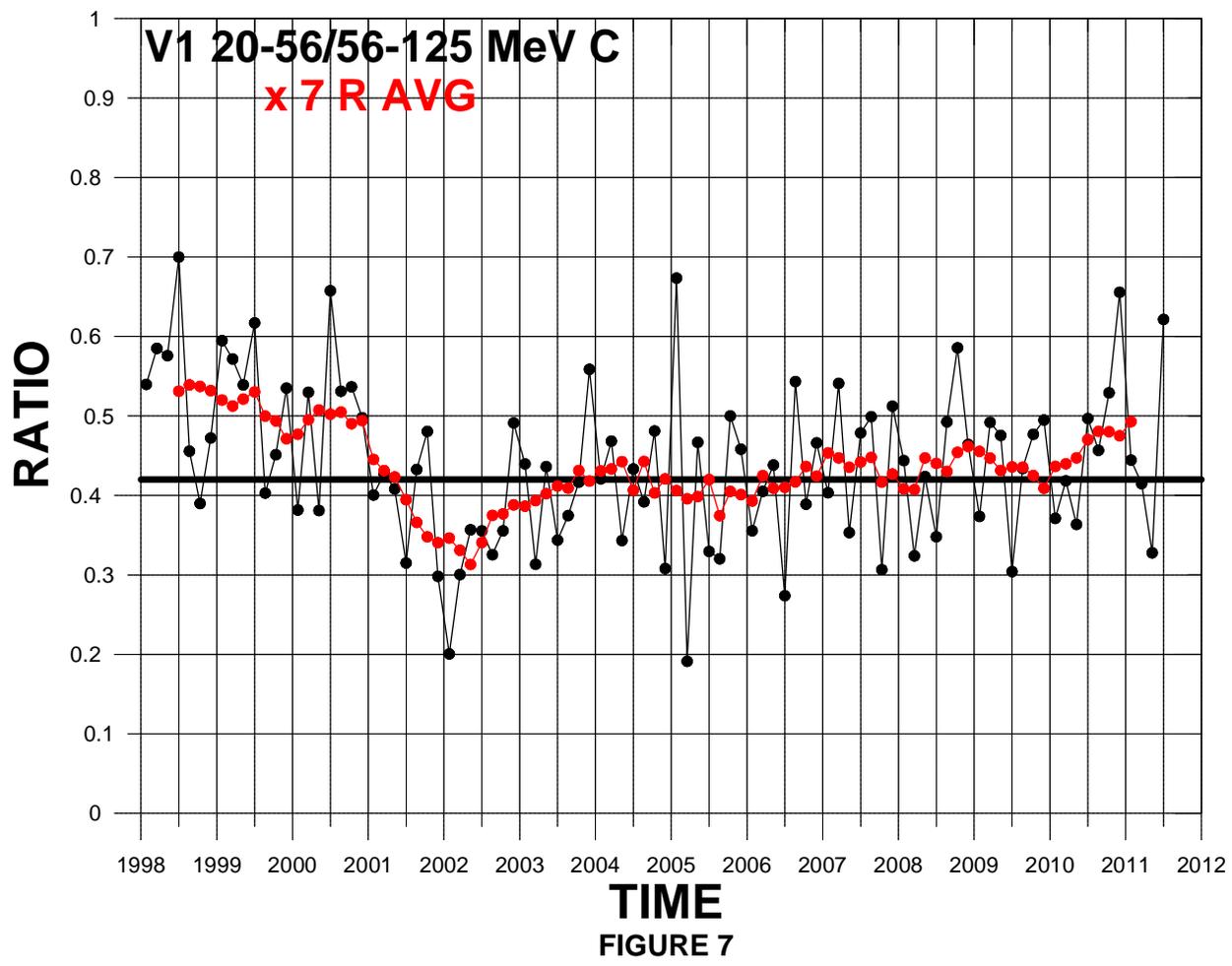

**FIGURE 7**